\documentclass[pdftex]{article}
\usepackage{spconf,amsmath,graphicx,bm,cite,hyperref}

\title{Singing voice synthesis based on frame-level sequence-to-sequence models considering vocal timing deviation}

\name{Miku Nishihara, Yukiya Hono, Kei Hashimoto, Yoshihiko Nankaku, and Keiichi Tokuda\thanks{This work was supported by JSPS KAKENHI Grant Number JP22H03614, CASIO SCIENCE PROMOTION FOUNDATION, and FOUNDATION OF PUBLIC INTEREST OF TATEMATSU.}}
\address{Department of Computer Science, Nagoya Institute of Technology, Nagoya, Japan}

\begin{document}
\ninept

\maketitle

\begin{abstract}
\vspace{-3pt}
This paper proposes singing voice synthesis (SVS) based on frame-level sequence-to-sequence models considering vocal timing deviation.
In SVS, it is essential to synchronize the timing of singing with temporal structures represented by scores, taking into account that there are differences between actual vocal timing and note start timing.
In many SVS systems including our previous work, phoneme-level score features are converted into frame-level ones on the basis of phoneme boundaries obtained by external aligners to take into account note durations of the musical scores and vocal timing deviations.
Therefore, the naturalness of the synthesized singing voices is affected by the aligner accuracy in conventional systems.
To alleviate this problem, we introduce an attention mechanism with frame-level features.
In the proposed system, the attention mechanism absorbs alignment errors in phoneme boundaries.
Additionally, we evaluate the system with pseudo-phoneme-boundaries defined by heuristic rules based on musical scores when there is no aligner.
The experimental results show the effectiveness of the proposed system.
\end{abstract}


\vspace{-3pt}
\section{Introduction}
\label{sec:intro}
\vspace{-5pt}
Statistical parametric singing voice synthesis (SVS)~\cite{oura2010recent,hono2021sinsy,hono2019singing,nishimura2016singing} has been developed along with the spread of machine learning.
This method statistically models not only acoustic features but also the temporal structures of singing voices because such a structure is highly dependent on the tempo and note durations of the musical piece.
Although the musical score information is input into the system, the actual vocal timing is usually different from the note start timing described in the musical score.
That is why modeling the temporal structure synchronized with the musical score has become an important task in SVS.

SVS systems based on deep neural networks (DNNs)~\cite{hono2021sinsy,hono2019singing,nishimura2016singing} that can generate natural singing have become commonly used.
The systems use DNNs to model the mapping function from score features, such as notes and phonemes, to acoustic features extracted from singing voices.
Generally, the lengths of phoneme-level score features used as input and that of frame-level acoustic features used as output are different.
In our previous work, Sinsy~\cite{hono2021sinsy}, phoneme-level score feature sequences are converted to frame-level ones on the basis of phoneme boundaries obtained by using pre-trained hidden semi-Markov models (HSMMs)~\cite{zen2007hidden} as an aligner for training DNNs.
During synthesis, the time-lag model and phoneme duration model determine the phoneme boundaries of the singing voice by taking into account vocal timing deviations related to the note start time given by the musical score.

Text-to-speech (TTS) synthesis systems based on a sequence-to-sequence (seq2seq) model~\cite{shen2018natural,ren2019fastspeech,ren2020fastspeech,li2019neural,ping2017deep,yu2020durian,kim2021conditional,cho2022mandarin,lim2022jets} have been reported to generate natural speech.
SVS systems based on a seq2seq model have also been proposed~\cite{gu2021bytesing,blaauw2020sequence,lee2019adversarially,lu2020xiaoicesing,chen2020hifisinger,wu2020adversarially,shi2021sequence} following seq2seq TTS.
However, it is difficult for seq2seq models to figure out the relationships of feature sequences with significantly different sequence lengths, such as a phoneme-level score feature sequence and a frame-level acoustic one all at once.
Furthermore, for SVS, it is difficult to predict acoustic features even for the same phoneme or the same singer because the temporal structure of the singing voice depends on the tempo and note duration of the song.
Therefore, in this paper, we focus on a framework used in many SVS systems including our previous work where phoneme-level score feature sequences are converted to frame-level ones in advance on the basis of phoneme boundaries.

We have developed an SVS system called Sinsy~\cite{hono2021sinsy}.
In this system, acoustic features and temporal structures of singing voices are modeled by independent DNNs.
This system can synthesize natural singing voices in accordance with the input musical scores without complicated operations like manual parameter adjustment.
However, in this system, the quality of the synthesized singing voice significantly depends on the accuracy of phoneme boundaries, i.e. alignments between phonemes and acoustic features, used in training.
Although phoneme boundaries used in training are usually estimated by external aligners such as pre-trained HSMMs, obtained phoneme boundaries often include alignment errors, which affect subsequent acoustic model training.

In this paper, to solve this problem, we introduce an attention mechanism into the acoustic model of Sinsy.
The attention mechanism captures the temporal structure of the training data and can absorb alignment errors in phoneme boundaries due to the aligner.
In addition, the attention mechanism enables high flexibility in weighting when converting features, which is expected to improve the sound quality.
We also introduce pseudo-phoneme-boundaries which are boundaries determined by heuristic rules.
Since no external model is used to estimate the pseudo-phoneme-boundaries, there is no need to use forced alignment.
We compare these methods in experiments and describe the relationship among the alignment accuracy in the proposed system.

\vspace{-3pt}
\section{DNN-based singing voice synthesis}
\label{sec:DNNsvs}
\vspace{-5pt}
Generally in TTS synthesis systems, a change of duration from the original speech is acceptable as long as it is not much different from the actual one.
However, if we synthesize singing voices regardless of the temporal structure, it may cause unusual sounds due to the unacceptable deviation of the beat between the singing voice and the accompaniment.
Therefore, it is essential to synchronize the timing of singing with temporal structures, such as tempo and note duration, represented in scores.

\begin{figure}[tb]
  \begin{center}
  \centerline{\includegraphics[width=0.98\hsize]{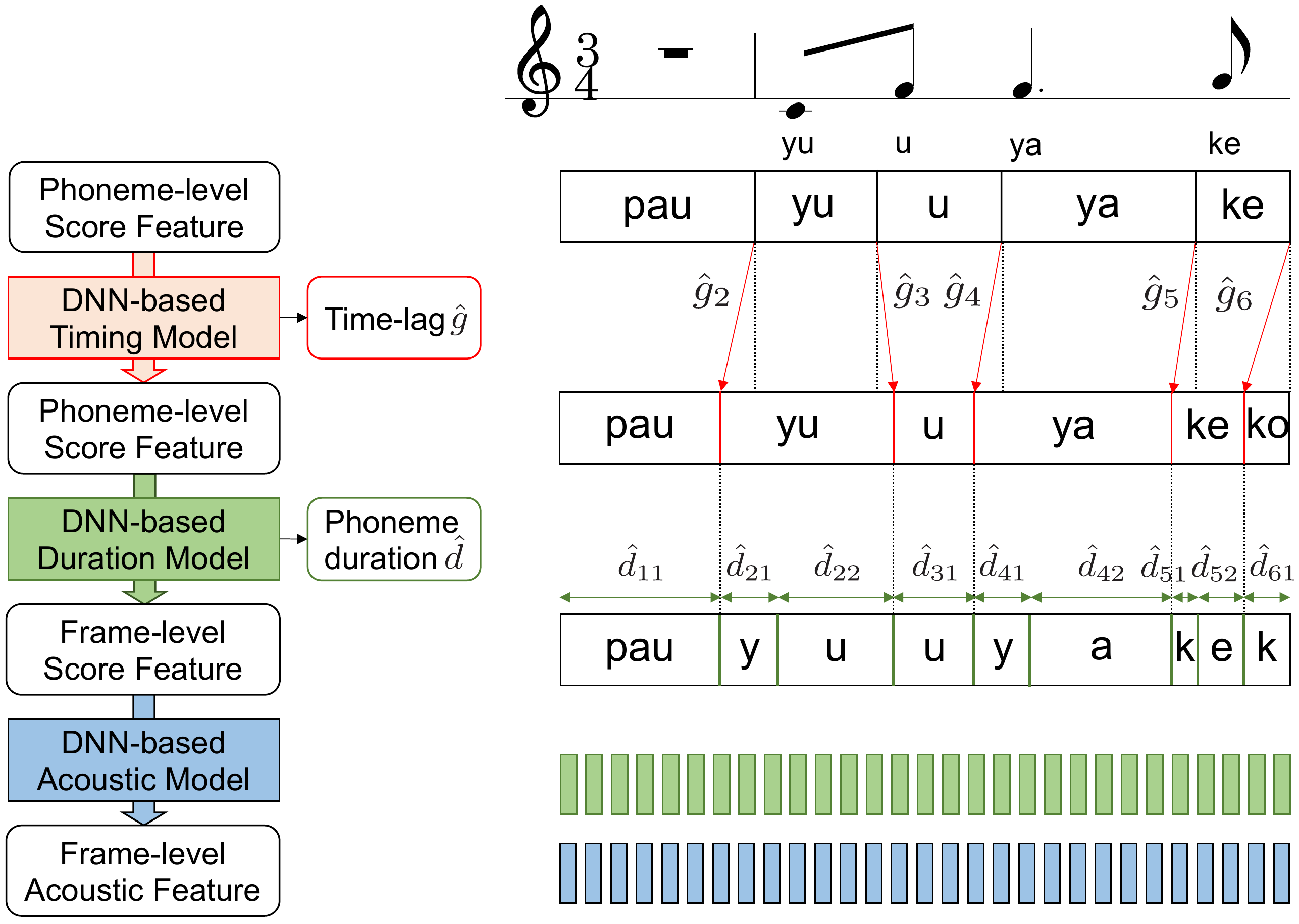}}
  \vspace{-32pt}
  \end{center}
  \caption{Outline of the DNN-based SVS system.}
  \label{fig:DNN}
  \vspace{-8pt}
\end{figure}

Figure~\ref{fig:DNN} shows the outline of the conventional DNN-based SVS system~\cite{hono2021sinsy}.
This system models acoustic features and temporal structures independently.
Acoustic features represent spectrum and fundamental frequency, and score features represent lyrics, note duration, and pitch described in musical scores.
In model training, phoneme boundaries are first estimated by aligners such as HSMMs.
Then, time-lag models, phoneme duration models, and acoustic models are trained by using DNNs, respectively.
In synthesis, phoneme-level score feature sequences without information of each phoneme durations are first converted to phoneme-level score feature ones by the time-lag models.
Then, frame-level feature sequences are obtained on the basis of the phoneme-level ones and phoneme boundaries determined by the phoneme duration models.
Finally, frame-level acoustic features are generated by inputting the obtained frame-level score features into the acoustic model.
DNN for acoustic models represent mapping functions from frame-level score features to frame-level acoustic ones.
Therefore, accurate phoneme boundaries are essential for training appropriate acoustic models and generating singing voices that are synchronized with the musical score.

There are also cases when vocal timing deviations occur because of intentional singing expressions or unique habits of the singer.
Because of this, the start position of the note determined from the musical score and the actual vocal timing are not always the same.
This deviation in vocal timing is defined as the difference between the start time of a note and that of the first phoneme of the note in this paper.
The note start timings and phoneme duration obtained from the scores deviate from the actual singing timing, which must be taken into account to match synthesized singing voice to the tempo of the song in the case of SVS.

The deviation estimated by the time-lag model is used for the estimation of note boundaries, and the phoneme duration model is used for the estimation of phoneme boundaries based on them.
Therefore, appropriate modeling of vocal timing deviations is one of the essential factors to generate a natural singing voice.

The sequence of note duration $\bm{L}$ and the sequence of deviations predicted by the DNN-based time-lag model $\hat{\bm{g}}$ are represented as follows:
\begin{align}
  \bm{L}&=(L_1, L_2, \ldots, L_N), 
  \label {eq:L}\\
  \hat{\bm{g}}&=(\hat{g}_1, \hat{g}_2, \ldots, \hat{g}_N),
  \label{eq:g}
\end{align}
where $N$ is the number of notes included in a song.
Note that it is always $\hat{g}_1=0$ because there is no timing deviations in the first note.
The $n$-th adjusted note duration $\hat{L}_n$ is obtained by
\begin{equation}
  \hat{L}_n= \left\{
    \begin{array}{ll}
      L_n - \hat{g}_n + \hat{g}_{n+1} & (n<N), \\
      L_n - \hat{g}_n & (n=N). 
    \end{array}
  \right.
\end{equation}
In this paper, the duration distribution of the $k$-th phoneme in the $n$-th note is represented by the Gaussian distribution with mean $\mu_{nk}$ and variance $\sigma_{nk}^2$.
The phoneme duration $\hat{d}_{nk}$ is calculated under the constraint of the note duration $\hat{L}_n$, which is obtained by taking into account the deviation in the vocal timing, as follows 
\begin{align}
  \hat{d}_{nk}&=\mu_{nk}+\rho_n \sigma_{nk}^2, 
  \label{eq:d-DNN} \\
  \rho_n&=\frac{\hat{L}_n-\sum_{k=1}^{K_n} \mu_{nk}}{\sum_{k=1}^{K_n} \sigma_{nk}^2},
  \label{eq:rho}
\end{align}
where $K_n$ is the number of phonemes in the $n$-th note.
Once we have calculated $\hat{d}_{nk}$, frame-level score features can be appropriately obtained by considering the vocal timing.

\vspace{-3pt}
\section{Proposed system}
\label{sec:frame2frame}
\begin{figure}[tb]
\begin{center}
  \centerline{\includegraphics[width=0.98\hsize]{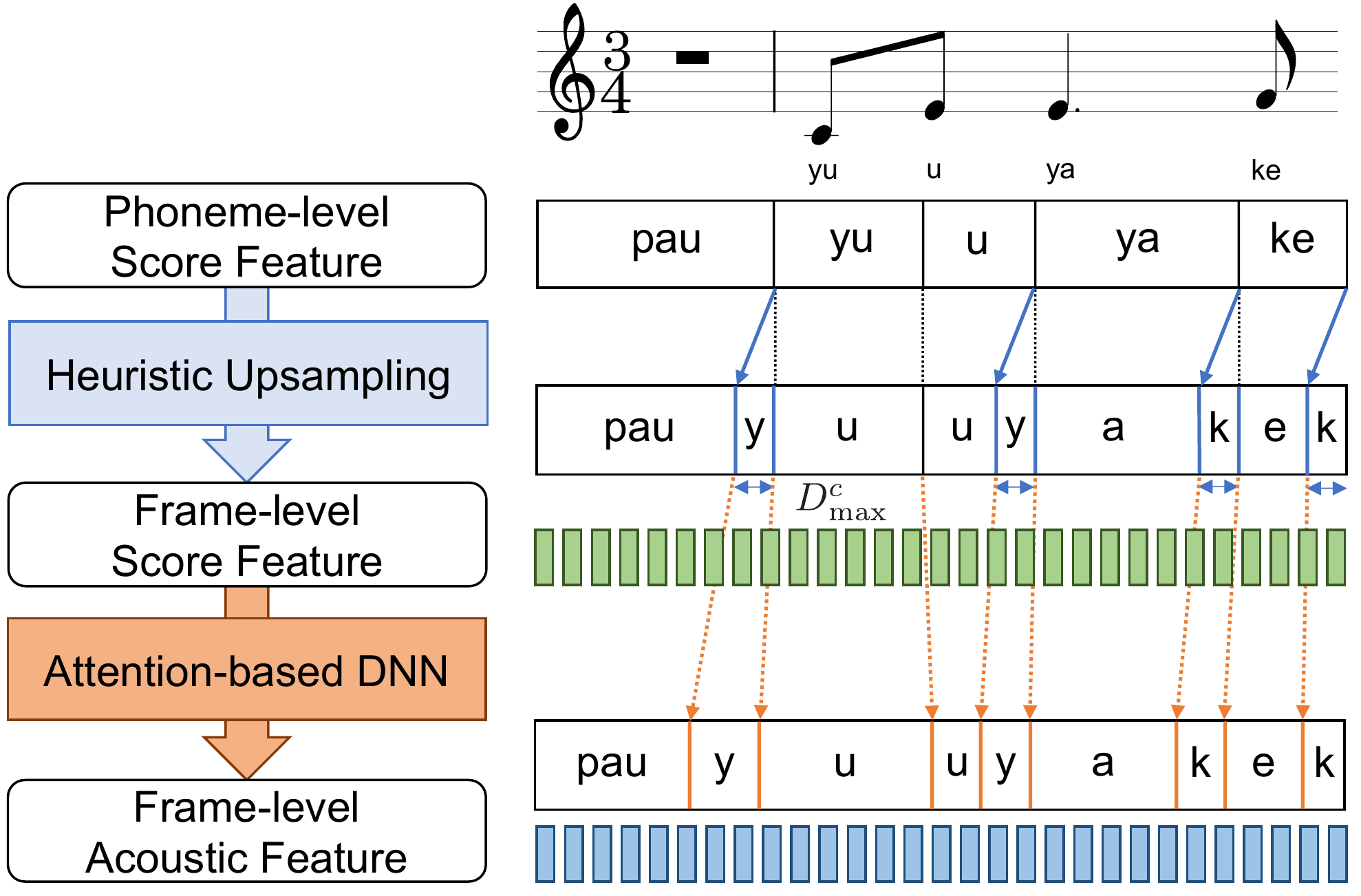}}
  \vspace{-30pt}
  \end{center}
  \caption{Outline of seq2seq SVS system based on pseudo-phoneme-boundaries.}
  \label{fig:pseudo}
  \vspace{-8pt}
\end{figure}

\vspace{-5pt}
In the conventional SVS systems, Sinsy~\cite{hono2021sinsy}, the alignment accuracy is easily affected so the alignment errors might be propagated.
To address this problem, we adapt frame-level seq2seq models based on an attention mechanism to our conventional SVS system to absorb the alignment errors.
The attention mechanism can absorb deviations between the actual vocal timing and estimated phoneme boundaries.
We also propose a method using pseudo-phoneme-boundaries that does not require forced alignment for estimating phoneme boundaries.
Figure~\ref{fig:pseudo} shows an outline of this system.

\vspace{-5pt}
\subsection{Model architecture}
\begin{figure}[tb]
  \begin{center}
  \includegraphics[width=\hsize]{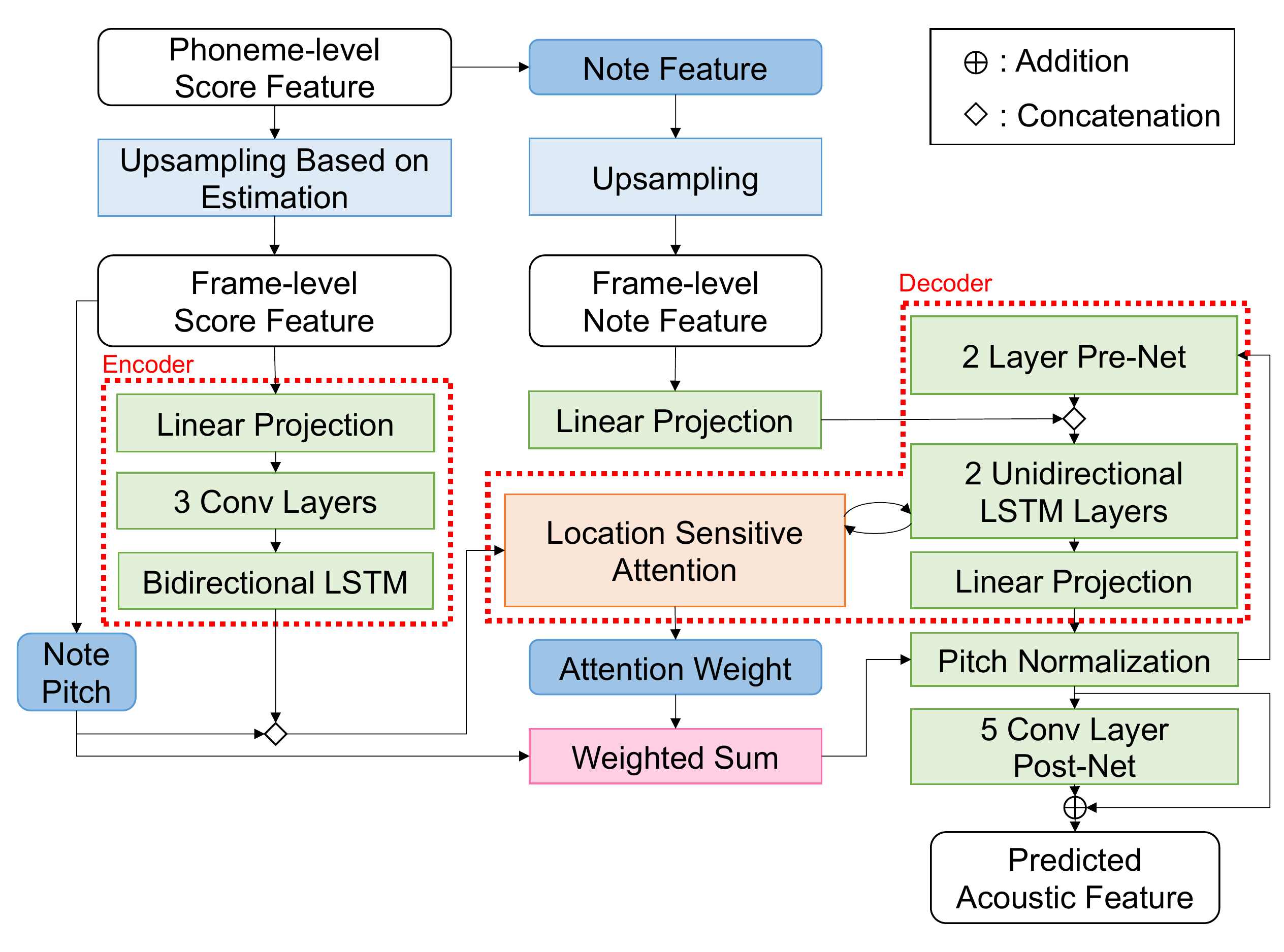}
  \vspace{-32pt}
  \end{center}
  \caption{Model architecture of seq2seq SVS system that is driven frame by frame.}
  \label{fig:taco2base}
  \vspace{-8pt}
\end{figure}

\vspace{-5pt}
\label{sec:architect}
Figure~\ref{fig:taco2base} shows the model architecture of SVS system based on frame-level seq2seq models.
The proposed system is based on the Tacotron~2~\cite{shen2018natural}.
Unlike Tacotron2, the attention mechanism in this paper is driven frame by frame.
The encoder uses as input the score features that have already been converted from phoneme-level to frame-level ones on the basis of previously determined phoneme boundaries.
The decoder outputs frame-level acoustic features by considering the deviation in vocal timing using the attention mechanism.

\vspace{-5pt}
\subsection{Embedding note features into the decoder}
\vspace{-5pt}
\label{sec:embedding}
In the proposed method, the vocal timing of the singing voice needs to be adjusted to the appropriate position on the basis of the musical score by the attention mechanism.
For this purpose, the decoder input is a vector concatenated from the output of the decoder Pre-Net and the extracted features of the target notes from the score features.
The frame-level note features are obtained by upsampling the note features in the phoneme-level score features in accordance with the note duration given by the musical score and concatenating it with the frame positioning information within the note.
Preliminary experiments confirmed that the note feature information used as an additional query enables the attention mechanism to adjust estimated phoneme boundaries into appropriate phoneme boundaries.

\vspace{-5pt}
\subsection{Pitch normalization}
\vspace{-5pt}
\label{sec:pitch}
Pitch normalization is performed as in the conventional SVS system~\cite{hono2021sinsy}, where the difference between the note pitch in the musical score and the logarithmic fundamental frequency (F0) of the singing voice is modeled by a DNN.
A note pitch sequence is obtained in accordance with the input musical score taking into account the actual vocal timing.
Pitch normalization requires a frame-level note pitch sequence, which is obtained by attention weighting of the note pitch sequence generated on the basis of the pre-estimated frame-level phoneme boundaries at each decoder step.

\vspace{-5pt}
\subsection{Training criteria}
\vspace{-5pt}
\label{sec:criteria}
The alignment obtained by the attention mechanism of the proposed method might follow the diagonal except near phoneme boundaries because the score feature sequences are converted to frame-level based on the estimated phoneme boundaries.
Therefore, we use a guided attention loss~\cite{tachibana2018efficiently} in addition to the summed mean squared error (MSE) from before and after the post-net.

\vspace{-5pt}
\subsection{Heuristic upsampling using the pseudo-phoneme-boundaries}
\vspace{-5pt}
\label{sec:pseudo}
We propose an aligner-free method using a heuristic rule that reflects our singing tendencies as appropriate phoneme boundaries are given by the attention mechanism.
As the heuristic rule for setting pseudo-phoneme-boundaries, this paper defines them in accordance with the rule of shifting the consonant of the first phoneme to the front under the constraint of maximum consonant duration $D_{\mathrm{max}}^c$.
The $k$-th phoneme duration in the $n$-th note $d_{nk}$ is given by
\begin{equation}
  d_{nk}=
  \begin{cases}
    d_n^c & \text{if~consonant}\\
    \dfrac{L_n - C_n \cdot d_n^c}{K_n - C_n} & \text{otherwise}
  \end{cases}
  \label{eq:d-pseudo}
\end{equation}
where $C_n$ is the number of consonants in the $n$-th note, and $d_n^c$ is
\begin{equation}
  d_n^c=\min (D_{\mathrm{max}}^c,~L_n/K_n)
  \label{eq:d-cons}
\end{equation}
In this paper, $D_{\mathrm{max}}^c$ is set to 30~frames.
As previously described, there is a deviation between the start timing of notes on the score and the actual vocal timing.
Since it is difficult to express the pitch of a note when it is a consonant, it tends to be pronounced a little earlier, so the vowel is assigned to the start position of the note.
The same approach is used in the time-lag model used in the conventional SVS system\cite{hono2021sinsy}.
Preliminary experiments also indicated that shifting consonants forward improves the naturalness of the synthesized singing voice.
In addition, the duration of consonants is not significantly affected by the note duration and is known to be shorter than that of vowels.
To take into account these tendencies, we adopted this heuristic rule, which shifts the first consonant of a note to the previous note position and sets the maximum consonant duration.

\vspace{-5pt}
\subsection{Relation to prior work}
\vspace{-5pt}
\label{sec:prior}
In Sinsy~\cite{hono2021sinsy}, the score features are converted from phoneme-level to frame-level ones on the basis of phoneme boundaries in advance to enable the DNN to train the correspondence between phoneme-level score features and frame-level acoustic ones.
During the training stage, the phoneme boundaries are obtained by forced alignment used the trained HSMMs, and during the synthesis stage, the boundaries are estimated explicitly from the vocal timing and phoneme duration models in accordance with Eq.~(\ref{eq:d-DNN}).
In the proposed system, the score features input to the seq2seq model are converted in advance from phoneme-level to frame-level ones on the basis of the phoneme boundaries like for Sinsy.
Furthermore, the seq2seq model models the relationship between the score and acoustic features.

Bytesing\cite{gu2021bytesing} is a seq2seq SVS system using score features converted into frame-level based on a duration model, which is similar to the proposed method.
However, the synthesized singing voices by Bytesing don't follow actual vocal timing because it does not take into account vocal timing deviations from the musical scores and they must be manually adjusted to mix with the musical accompaniment.
In contrast, the proposed system can synthesize voices that are synchronized with the temporal structure of the score because it takes into account the vocal timing deviation when estimating phoneme boundaries and is absorbed by the attention mechanism.
On the other hand, strategies that reduce the dependence on alignment have also been proposed.
The authors of \cite{blaauw2020sequence} produces frame-level score features with phoneme boundaries estimated by a simple duration model on the basis of average phoneme durations and absorbs vocal timing deviations by using a seq2seq model with self-attention.
This requires manual correction to construct an appropriate duration model.
Incidentally, \cite{lee2019adversarially} does not use any external model but uses heuristic rules to generate frame-level score features like our proposed method.
When multiple phonemes are assigned to a single note, \cite{lee2019adversarially} assumes insufficiently, whereas we expect our system to maintain naturalness for the variable number of consonants and long tones since it constrains the maximum consonant duration.

\vspace{-3pt}
\section{Experiments}
\label{sec:exp}
\vspace{-5pt}

\subsection{Experimental conditions}
\vspace{-5pt}
\label{sec:conditions}
We conducted experiments using 70 Japanese children's songs performed by a single female singer: 60 songs were used for training, and the rest were used for evaluation.
Singing voice signals were sampled at 48 kHz and each sample was quantized by 16 bits.
The acoustic feature consisted of 50-dimensional mel-cepstral coefficients extracted by WORLD~\cite{morise2016world}, a continuous log F0 value, 25-dimensional analysis aperiodicity measures, 1-dimensional vibrato component, and a voiced/unvoiced binary flag.
The difference between the original log F0 and the median-smoothed one was used as the vibrato component.
The acoustic features were extracted with a 5-ms frame shift.
The frame-level score feature for the input of the encoder was a 272-dimensional feature vector consisting of the context of the current phoneme and note information, frame position in the current phoneme, and the phoneme duration.
The frame-level note feature for the input of the decoder was a 92-dimensional feature vector consisting of the context of the current note, the frame position in the current note, and the note duration.
The reduction factor was set to 3 for all methods.
All methods were combined with the pre-trained PeriodNet~\cite{hono2021periodnet}, a non-AR neural vocoder with a parallel structure, to generate waveforms from predicted acoustic features.
Five-state, left-to-right, no-skip HSMMs were used to obtain the alignment of the score and acoustic features.
The following four systems were compared.
\setlength{\leftmargini}{9pt}{
\begin{description}
  \item[\textbf{fal w/o att}] In the training stage, phoneme boundaries obtained by forced alignment using trained HSMMs are used, while in the synthesis stage, they were explicitly estimated using the time-lag model and phoneme duration model in accordance with Eq.~(\ref{eq:d-DNN}).
  This system architecture is similar to the conventional SVS system~\cite{hono2021sinsy} that uses multiple DNNs, except that the acoustic model structure has been modified on the basis of Tacotron~2-like encoder-decoder model without the attention mechanism.
  \item[\textbf{fal w/ att}] The method to define phoneme boundaries is the same as that of \textbf{fal w/o att} and the attention mechanism is used in this system. 
  \item[\textbf{model w/ att}] In both the training and synthesis stages, the boundaries were explicitly estimated using the time-lag model and phoneme duration model in accordance with Eq.~(\ref{eq:d-DNN}).
  \item[\textbf{pseudo w/ att}] In both the training and synthesis stages, the first consonants in each note were deliberately shifted and then the pseudo-phoneme-boundaries were calculated in accordance with Eq.~(\ref{eq:d-pseudo}).
  \end{description}
}
The systems, \textbf{fal w/ att}, \textbf{model w/ att}, and \textbf{pseudo w/ att}, are based on frame-level seq2seq models.
The difference between them is the phoneme boundaries for extracting frame-level score features.
As comparison methods, \textbf{model w/o att} and \textbf{pseudo w/o att} were also considered.
However, we excluded them from the experiments because they generated unnatural singing voice samples such as missing phonemes due to failing to absorb the mismatch of phoneme boundaries without attention mechanisms.

\vspace{-5pt}
\subsection{Subjective evaluation}
\vspace{-5pt}
\label{sec:eval}
We performed two mean opinion score (MOS) tests to evaluate the naturalness of the synthesized singing voices.
Each of the 14 native Japanese-speaking participants evaluated 15 phrases averaging about 10 seconds randomly selected from the test songs.
In the first experiment, we asked the participants to evaluate the quality of the synthesized singing voices, focusing on the naturalness of the sound quality and pitch.
In the second experiment, a click sound generated on the basis of the tempo of the score was mixed with the synthesized singing voices and the naturalness of the vocal timing was evaluated.

\begin{figure}[htb]
\vspace{5pt}  
\begin{minipage}[b]{.48\linewidth}
  \centering
  \centerline{\includegraphics[width=\hsize]{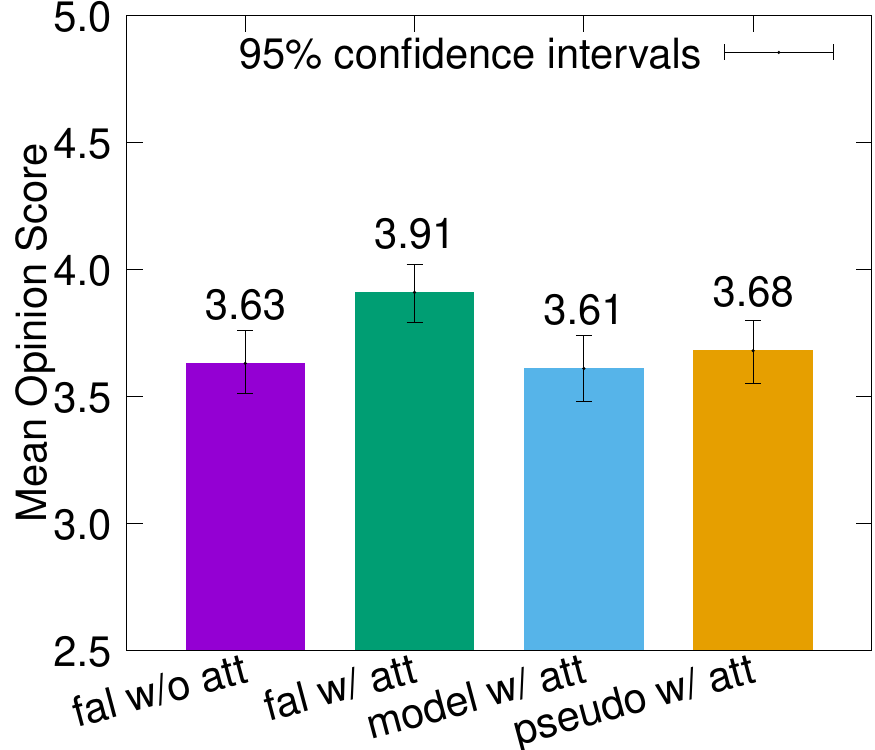}}
  \centerline{(a) The score for the sound quality.}
\end{minipage}
\hfill
\begin{minipage}[b]{0.48\linewidth}
  \centering
  \centerline{\includegraphics[width=\hsize]{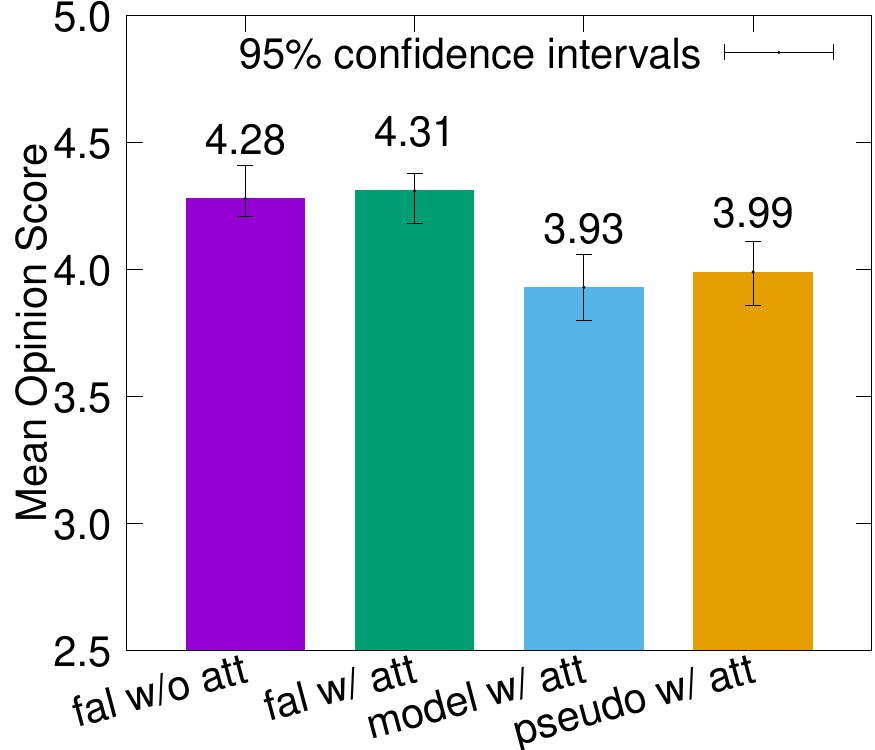}}
  \centerline{(b) The score for the vocal timing.}
\end{minipage}
\vspace{-5pt}
\caption{Mean opinion scores.}
\label{fig:mos}
\vspace{-8pt}
\end{figure}

The MOS results are plotted in Fig.~\ref{fig:mos}.\footnotemark \footnotetext{Audio samples: \url{https://www.sp.nitech.ac.jp/~mkring/demo/frame-seq2seq-svs/}}
Compared with \textbf{fal w/o att}, \textbf{fal w/ att} which is the proposed system greatly improved the naturalness score.
These results suggest that the use of an attention mechanism improves sound quality compared with the conventional DNN-based SVS method.
The scores for vocal timing of \textbf{fal w/ att} and \textbf{fal w/o att} were almost the same.

In \textbf{fal w/ att}, the time-lag and phoneme duration models used to estimate phoneme boundaries during synthesis were the same as those used in \textbf{model w/ att} to estimate boundaries during both training and synthesis.
However, for \textbf{fal w/ att}, the phoneme boundaries are obtained from the forced alignment obtained from the actual singing voice data during training.
In addition, \textbf{model w/ att} generated voices that had a tendency to be delayed and unstable regarding the vocal timing.
When we checked the phoneme boundaries estimated by the time-lag and phoneme duration models, the start timing of each phoneme was later and more fluctuating than other methods.
The results comparing \textbf{fal w/ att} and \textbf{model w/ att} also indicate the amount of vocal timing deviation and the alignment errors that can be absorbed by the attention mechanism are limited because the phoneme boundaries estimated by the time-lag and phoneme duration models for training data were less accurate than the forced alignment.
Therefore, the accuracy of phoneme boundary estimation during training has a strong effect on naturalness.

\textbf{pseudo w/ att} which is another proposed system and \textbf{model w/ att} showed similar scores in both tests.
In addition, these two methods showed synthesized sound quality comparable to the baseline system, \textbf{fal w/o att}.
This result indicates that a natural singing voice can be generated without any external models, and that appropriate vocal timing modeling can be achieved from pseudo-phoneme-boundaries based on the musical score.
The phoneme boundaries used in \textbf{pseudo w/ att} and \textbf{model w/ att} are not consistent with the actual vocal timing of the singing data.
Such deviations are appropriately modeled by the attention mechanism in our proposed system.
Although the timing is not as good as the conventional method, the results are sufficiently practical.

\vspace{-3pt}
\section{Conclusion}
\label{sec:conclusion}
\vspace{-5pt}
In this paper, we proposed frame-level seq2seq models considering vocal timing deviation to our conventional DNN-based SVS system, Sinsy.
We also proposed the method using pseudo-phoneme-boundaries that do not require external aligners for converting the musical score features from phoneme-level to frame-level.
Experimental results show that the proposed method improves the naturalness of the synthesized singing voice from the conventional method.
In addition, the naturalness score of the proposed method with pseudo-phoneme-boundaries was comparable to those of the conventional method with high-accuracy aligners.
Future work includes to compare our method with existing ones and to improve the accuracy of vocal timing estimation by the attention mechanism.

\vfill\pagebreak

\bibliographystyle{IEEEbib}
\bibliography{refer}

\end{document}